\documentclass{article}

\usepackage{PRIMEarxiv}

\usepackage[utf8]{inputenc} 
\usepackage[T1]{fontenc}    
\usepackage{hyperref}       
\usepackage{url}            
\usepackage{booktabs}       
\usepackage{amsfonts}       
\usepackage{nicefrac}       
\usepackage{microtype}      
\usepackage{amsmath}        
\usepackage{lipsum}
\usepackage{fancyhdr}       
\usepackage{graphicx}       
\usepackage{cleveref}
\crefname{equation}{Eq}{Eqs} 
\graphicspath{{media/}}     

\title{Triplet Envelope Functions for increasing machine learning interatomic potential efficiency and stability
}

\author{
  Emil Annevelink, Varun Shankar \\
  Physics Inverted Materials, Inc. \\
}

\begin{document}
\maketitle

\begin{abstract}
Central to interatomic potential efficiency is the radial envelope function that enables linear scaling with computational cost by defining a local neighborhood of atoms. 
This has enabled machine learning interatomic potentials (MLIPs) to revolutionize materials science over the past decade by providing density functional theory (DFT) accuracy with linear scaling computational cost in molecular dynamics workflows.
However, MLIPs still have a relatively high computational cost compared to empirical interatomic potentials preventing MLIP's from transforming most molecular dynamics workflows. 
A central issue is that MLIPs use relatively large cutoff radii, converging to 6 \AA\ over the last few years.
The large cutoffs prioritize accuracy of any material over efficiency in any particular region of phase space, capturing dispersion effects and low density materials at the expense of increased computational cost in higher density materials. 
Past work has aimed to address this with KNN graph sparsification, which, while significantly reducing cost, has the drawback of breaking energy conservation.
In this work, we propose higher-order envelope functions that prune local atomic neighborhoods through physically inspired geometric functions to provide the memory and efficiency benefits of KNN graph sparsification while eliminating non-conservative energy dynamics.
Through numerical experiments on solids and liquids with 5-8 \AA\ cutoffs, we show that triplet envelope functions complement radial envelope functions by doubling training and inference speed, tripling memory efficiency, and increasing simulation stability while not impacting accuracy or data efficiency for the most common 6\AA\ cutoff.
Moreover, experiments with 8 \AA\ radial cutoffs show triplet envelope functions create a pathway to larger cutoff radii for efficiently and accurately modeling open structures with large interatomic distances, showing a promising new direction for engineering MLIP efficiency.

\end{abstract}


\section{Introduction}
Machine learning interatomic potentials (MLIPs) are a new paradigm in computational materials science due to their reduced cost compared to density functional theory (DFT) calculations \cite{lysogorskiy2025grace}.
General purpose models trained on expansive DFT datasets now regularly provide quantitative accuracy across the periodic table \cite{ChipsffWines2025,MatBenchRiebesell2025, wood2025umafamilyuniversalmodels}.
As a replacement for ab-initio molecular dynamics, this is transformative, but the aspiration for MLIPs is to also replace empirical interatomic potentials.
However, current general purpose models are still orders of magnitude more costly than empirical interatomic potentials, and using MLIPs to probe the thermodynamics and kinetics of materials across nanosecond timescales is still prohibitively expensive \cite{hybridnnWen_2019}. 
These barriers have prevented them from being widely adopted across the computational materials science community, especially in industrial research settings.

Locality is the primary method MLIPs have reduced the cost of simulations with respect to DFT.
A local neighborhood is defined by a radial cutoff $R_c$ that limits the receptive field of MLIPs.
The number of atoms in the local environment determines how much information the MLIP processes and correspondingly its cost.
Importantly, locality makes interatomic potentials scale linearly with the system size for accessing much larger simulations than DFT.
However, since the number of atoms within the neighborhood increases cubically with the cutoff radius, increasing the cutoff has a significant impact on the model cost.

The cost needs to be balanced with accuracy. 
Reducing the radial cutoff generally comes at the expense of accuracy as the physics an MLIP needs to capture has to be within its receptive field.
For isolated small molecules, a cutoff radius of 2-3\AA\ is sufficient to model covalent bonding.
For most metals, a cutoff radius of 4-5 \AA\ is generally sufficient.
It is only when open inorganic structures or dispersion effects are modeled that cutoffs of 6 \AA\ or more are needed.

Before general purpose MLIPs gained popularity, the radial cutoff was a hyperparameter that balanced the accuracy and cost of MLIPs for the specific physics being modeled \cite{Sch_tt_2017, Lubbers_2018, Unke_2019}. 
To be accurate across the entire periodic table, MLIPs have adopted a uniform cutoff of 6 \AA\ that ensures accuracy across a wide range of materials \cite{batatia2025foundationmodelatomisticmaterials, fu2025learningsmoothexpressiveinteratomic, lysogorskiy2025grace}.
The uniform cutoff radius coupled with the variability in structure results in the number of neighbors being vastly different for different structures. 
For instance, a cutoff distance of 5 \AA\ in diamond creates a graph where each carbon atom has 5 nearest neighbor shells in its neighbor list.
Compare this to barium, where a 5 \AA\ cutoff only spans a single nearest neighbor shell. 
This asymmetry is what requires the large 6 \AA\ cutoff in general purpose models but also leads to increased costs for evaluating materials that could otherwise be treated with smaller cutoffs.

Previous attempts at addressing the large number of neighbors has been through sparsifying and pruning the neighbors for each atom \cite{gasteiger2024gemnetuniversaldirectionalgraph, radicalai2025egip}. 
The simplest is K nearest neighbor (KNN) sparsification, where the edges are sparsified by sorting them according to length, keeping only the K nearest neighbor edges \cite{rhodes2025orbv3atomisticsimulationscale}.
Each of these previous approaches have shown the promising gains of graph sparsification in terms of memory and cost improvements.
However, a primary drawback of each approach is that they break energy conservation.
Energy conservation is broken as neighbors are removed before their energy contribution smoothly goes to zero from the radial envelope function.
As has been shown previously \cite{bigi2025darkforcesassessingnonconservative}, removing the energy conservation of MLIPs can have significant consequences on the thermodynamic observables from molecular dynamics.
Because molecular dynamics is the primary use case of next generation low-cost MLIPs, previous sparsification methods are untenable.
Moreover, the isotropic nature of KNN sparsification can lead to removing critical anisotropic features.
For example, at solid-liquid interfaces, where the density of materials changes.

Despite a trend to remove constraints on MLIPs, incorporating physical priors has repeatedly proven an effective tool at improving MLIP performance \cite{gasteiger_dimenet_2020, nequipBatzner2022, fu2025learningsmoothexpressiveinteratomic}.
In particular, we aim to achieve the same benefits of graph sparsification -- reduced cost and memory -- without the drawbacks -- loss of energy conservation and naive pruning.
To this end, we investigate triplet envelope functions as principled methods for graph sparsification, drawing heavily on functionals from empirical interatomic potentials.
For example, in the modified embedded atom method (MEAM), the local electron density is constructed as a summation of contributions from atoms in a local environment \cite{meambaskes1992,2nnmeamLee2000, 2nnmeamlee2001}.
In particular, the MEAM potential improved how the local density was calculated by screening the contributions of atoms in a neighborhood, where the screening function distinguishes atoms in the first and second nearest neighbor shells and monotonically decreases due to angular overlap between triplets of atoms.

In this work, we investigate the feasibility of the MEAM screening function as a prototype higher-order envelope function.
Triplet envelope functions are a higher order, three-body complement to two-body radial envelope functions. 
The triplet envelope function defines an angular dependent method for reducing the interaction between atom pairs. 
When the envelope function goes to zero, we sparsify the neighbor lists of atoms reducing the cost of model inference without affecting energy conservation. 
Through numerical experiments, we show that triplet envelope functions have a similar effect as KNN sparsification by significantly reducing the number of edges to create a nearly uniform number of edges for different materials.
Moreover, we show how triplet envelope functions improve on radial envelope functions to reduce training and inference cost, reducing memory pressure, and increasing simulation stability.

\section{Methodology}
\label{sec:headings}

\subsection{Radial Cutoff and Envelope Functions}
Machine learning interatomic potentials embed pairwise edge information using radial basis functions.
The basis functions range from Gaussians, to Chebyshev \cite{acedrautz2019} or Bessel \cite{gasteiger_dimenet_2020} polynomials.
These basis functions are modulated with envelope functions that ensure the basis goes to zero as $ r \rightarrow R_c$.
Mathematically, the edge embedding is defined as
\begin{equation}
    e(r_{ij}) = b(r_{ij})f(r_{ij}) \hspace{0.01in},
\label{eq:radialembedding}
\end{equation}
where $e$ represents the edge embedding, $b$ the basis function, and $f$ the envelope function.
While the choice of envelope function is only constrained by the requirement it is zero when $r_{ij} = R_c$, it has been shown that using envelope functions that are twice continuously differentiable at the cutoff is beneficial because they ensure smooth Hessians and thus reliable geometry optimization and energy conservation during simulations due to a smooth potential energy surface \cite{gasteiger_dimenet_2020}. 
Common choices for envelope functions are therefore high order polynomials \cite{gasteiger_dimenet_2020} or periodic functions \cite{acedrautz2019}.

\subsection{Triplet-Based Envelope Function}
To extend the radial envelope function to higher order features, we extend \cref{eq:radialembedding} to include triplet interactions according to  
\begin{equation}
    f_t(r_{ij}) = \prod_{k\in N_i \cup N_j}{f_{ijk}} \hspace{0.01in},
    \label{eq:tripletenvelope}
\end{equation}
where $f_t$ is the triplet envelope function, k is an index for all atoms in the radial neighborhood of i, and $f_{ijk}$ is a three-body term.

The functional form of the triplet envelope function defined in \cref{eq:tripletenvelope} is very similar to how bond-order potentials such as Tersoff \cite{tersoff_1988} and AIREBO \cite{Stuart2000ARP} treat local environments and how screening is defined in MEAM interatomic potentials.
Specifically in the MEAM framework, the influence of atom $j$ on atom $i$ is determined by how much a third atom $k$ screens their interaction.
The total influence is determined by a screening function $S_{ij}$ that is the product of the screening from all atoms $k$ in both neighborhoods of atoms $i$ and $j$ according to

\begin{equation}
S_{ij} = \prod_{k\in N_i \cup N_j}{S_{ijk}} \hspace{0.01in}.
\label{eq:Sij}
\end{equation}
Although developed for metallic systems, the MEAM screening function is purely geometric and smoothly identifies neighbors in the local environment that are expected to contribute to the local electron density seen by atom $i$.
In this way, it is a generally applicable physical principle from quantum mechanics that can be applied to any interatomic interaction.

\begin{figure}
    \centering
    \includegraphics[width=0.8\linewidth]{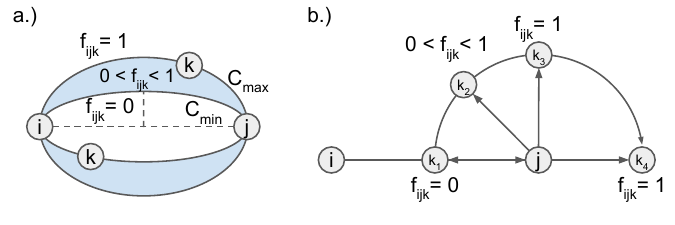}
    \caption{Geometric representation of triplet screening. (a) Schematic of the ellipse-based screening geometry, where the position of atom $k$ relative to atoms $i$ and $j$ determines its contribution to the envelope function $f_{ijk}$. (b) Progression of the screening function $S_{ijk}$ across representative three-body configurations, illustrating the transition from fully screened ($S_{ijk}=0$) to unscreened ($S_{ijk}=1$) states. }
    \label{fig:screening}
\end{figure}

Following the MEAM framework, we adopt a functional form of $f_{ijk}$ that mirrors the geometric screening term $S_{ijk}$.
The functional form of $S_{ijk}$ governs the triplet screening by finding the ellipse created by the three atoms $i-j-k$.
As shown in Figure \ref{fig:screening}, one of the ellipse axes is defined as the distance $r_{ij}$, and the ratio of the ellipse axes is defined as $C_{ijk}$, which can be found using the remaining two interatomic distances $r_{ik}$ and $r_{jk}$

\begin{equation}
    C_{ijk} = 1 + 2\frac{r_{ij}^2r_{ik}^2+r_{ij}^2r_{jk}^2 - r_{ij}^4}{r_{ij}^4-(r_{ik}^2-r_{jk}^2)^2} \hspace{.01 in}.
\label{eq:cijk}
\end{equation}
To find the magnitude atom $k$ screens the interaction between atoms $i$ and $j$, the axis ratio $C_{ijk}$ is compared to minimum and maximum values $C_{min}$ and $C_{max}$ to yield the following equation for $S_{ijk}$

\begin{equation}
    S_{ijk} = f_c \left( \frac{C_{ijk} - C_{min}}{C_{max}-C_{min}} \right) \hspace{.01in},
    \label{eq:Sijk}
\end{equation}
where the truncation function smoothly varies from $0\rightarrow1$ and is twice differentiable and defined as 

\begin{equation}
    f_c(x) =
    \begin{cases}
    1, & x \ge 1 \\
    \left[ 1 - (1 - x)^4 \right]^2, & 0 < x < 1 \\
    0, & x \le 0
    \end{cases}
    \label{eq:fc}
\end{equation}

A graphical representation of \cref{eq:cijk,eq:Sijk,eq:fc} is visualized in Figure \ref{fig:screening}.
Using the ellipsoidal definition, as a third atom $k$ is closer to bisecting atoms $i$ and $j$, the strength of the $i-j$ interaction decreases.
Similarly, as a third atom moves away from the connecting line, $f_{ijk}$ increases, reducing the effect atom $k$ has on the interaction.
The bounds of where a triplet begins screening are defined by $C_{min}$ and $C_{max}$.
The more critical of these is $C_{min}$, which defines the ellipse where the $i-j$ interaction is completely screened.
When any $C_{ijk}$ is below $C_{min}$, $S_{ij}$ is zero.

The triplet screening is incorporated into the edge embedding through
\begin{equation}
    e(r_{ij}) = b(r_{ij})f_r(r_{ij})f_t(r_{ij}) \hspace{0.01in},
\label{eq:tripletembedding}
\end{equation}
where $f_r$ and $f_t$ represent the radial and triplet envelope functions respectively.
By using the definition of $S_{ij}$ for $f_t(r_{ij})$, the triplet envelope function can be used directly within common MLIP implementations. The choice of $C_{min}$ and $C_{max}$ become model hyperparameters akin to the radial cutoff, which can be determined through hyperparameter optimization.

\subsection{Logistic Filtering of High Frequency Modes}
It is important to consider that the MEAM interatomic potential was originally developed for metals with nearest neighbor distances $r_{min} > 2$\AA. 
However, complex materials and organic molecules in particular routinely have interatomic spacings near 1\AA.
These small distances can introduce high-frequency modes in the screening function that must be filtered out to maintain stability with reasonable timesteps.
For instance, for van-der-Waals interactions between two molecules, $R_{ij}$ is typically 2-3x larger than intramolecular distances $R_{jk}$. 
Due to the acute triplet angle caused by the anisotropic distances, atom $k$ would completely screen the interaction if $R_{jk}<R_{ij}$ and provide no screening if $R_{jk}>R_{ij}$. 
This is seen in Figure \ref{fig:screening}b in the $k_3$ location, where for small values of $R_{jk}$, the intermediate regime $k_2$ is very small.

We introduce a logistic filter to remove the high frequency modes by creating an exclusion region for triplets.
A sigmoid function $\sigma(x) = \frac{1}{1+\exp(-x)}$ on each of the triplet distances $r_{ik}$ and $r_{jk}$ create a minimum distance that atoms interact with each other. The two sigmoid functions are combined according to

\begin{equation}
    F_{ijk,LP} = 1 - \sigma(k(r_{ik} - r_0)) \cdot \sigma(k(r_{jk} - r_0)),
\end{equation}
where $k$ is the sigmoid width and $r_0$ is the sigmoid location. The low pass filter $F_{ijk,LP}$ is used to create a new low-pass filtered $S_{ijk,LP}$ through

\begin{equation}
    S_{ijk,LP} = S_{ijk} + (1-S_{ijk})F_{ijk,LP},
\end{equation}
which modulates the screening such that $S_{ijk,LP}=1$ when $F_{ijk,LP}=1$ and $S_{ijk,LP}=S_{ijk}$ when $F_{ijk,LP}=0$.
This is used in \cref{eq:Sij}, where the filtered $S_{ijk,LP}$ replaces the triplet screening $S_{ijk}$ to have a low-pass filtered $S_{ij}$.
By incorporating the low-pass filters, the screening from atoms with distances $r_{ik}\approx r_0$ and $r_{jk} \approx r_0$ is minimized limiting the high frequency modes.
$k$ and $r_0$ become additional hyperparameters which can be used to reduce the high frequency modes such as those from rotations of small molecules.
This means setting a limit on the minimal size of the transition region $k_2$ in Figure \ref{fig:screening}b, which can be done for most atomic systems by ensuring setting the sigmoid exclusion zone $>1$\AA.
The screening function could further be improved by non-dimensionalizing the hyperparameters to create a more general description.

\subsection{Implementation}
The triplet envelope function is implemented in PyTorch and tested in PHIN’s active learning framework for running molecular dynamics simulations. 
Similar to two-body envelope functions, the triplet envelope functions are used to sparsify the atomic graphs before gradients are tracked to increase MLIP efficiency.
During training, graph sparsification occurs alongside traditional preprocessing steps that construct the atomic graph. 
Because the triplet screening computation scales as $\mathcal{O}(n^3)$, where n are the neighbors in the radial cutoff, we precompute the edge sparsification to maximize the efficiency gained from reducing the edges.
For inference, however, graph sparsification is done on-the-fly at each timestep making the sparsification overhead a major liability.
We tested GPU and multi-threaded CPU preprocessing and found that the multi-threaded implementation in LAMMPS was faster by parallelizing over the outermost neighbor loop, although it is likely that dedicated CUDA kernels for computing triplets may be faster still.
For all experiments, the triplet envelope function uses $C_{min}=0.3$ and $C_{max}=1.5$.  
Compared to $C_{min}$ and $C_{max}$ from fitted MEAM values, these are significantly smaller, making them conservative in what interactions to geometrically screen.
We found that $r_0=1.5$\AA\ and $k=5$\AA$^{-1}$ are reasonable defaults that can be applied to the chemical systems that were tested.

\subsection{Training datasets}

Training datasets for silicon and water are generated using PHIN's active learning software.
The silicon dataset is generated to converge a 50ps NVT simulation of 64 silicon at 3000K. The simulation starts from a cubic diamond lattice and melts after 5ps such that both crystalline and amorphous, melted structures are contained in the dataset.
The converged silicon dataset contains 21 structures representing 1344 local atomic environments.

The water dataset is generated to converge a 10ps NVT simulation of 20 water molecules at 300K. The simulation starts from a randomly initialized structure created using Packmol \cite{Martinez2009}.
The 10ps simulation is long enough for the water molecules to rotate ensuring that the dataset contains both intramolecular and intermolecular complexity as the bonds vibrate and the molecules change orientations with respect to one another.
The converged water dataset contains 75 structures representing 4500 local atomic environments.

\section{Results}
\label{sec:results}

\begin{figure}
  \centering
  \includegraphics[width=0.7\linewidth]{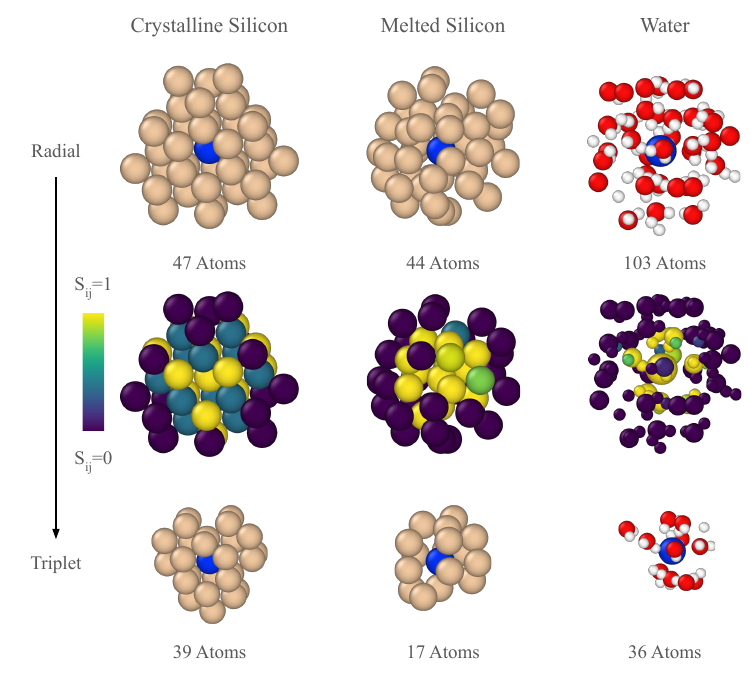}
  \caption{Comparison of neighborhoods defined by radial and triplet envelope functions for a radial cutoff of 6\AA. The top two rows depict the local neighborhood of a central atom (blue), where the top row is colored by chemical identity and the middle row by the triplet screening parameter. The bottom row depicts the local environments after pruning with the triplet envelope function.}
  \label{fig:neighborhoods}
\end{figure}

The triplet envelope function is tested on two benchmark systems, crystalline/melted silicon and liquid water. 
The different local environments that each system sees is visualized in Figure \ref{fig:neighborhoods}. 
For all systems, the radial cutoff has a much larger number of edges than the triplet envelope function showcasing the sparser atomic graphs created by the triplet envelope function.
An interesting feature of the three sparsified environments is the local structure captured in each neighborhood.
While the radial cutoff is spherical as expected for all three environments, the triplet envelope function creates structure specific local environments.
For crystalline silicon, the local environment resembles the crystalline diamond cubic structure of silicon.
As the silicon melts and the local structure homogenizes in the liquid phase, the neighborhood identified by the triplet screening is more spherical although with a smaller radius.
However, even though water is also a liquid, the triplet function picks up not just a smaller radius but also the local structure of the mixed covalent and van-der-Waals structure leading to more of an ellipsoidal shape.
The ability of the triplet envelope function to distinguish between each of the different structures is a key to the formalism and shows how it is able to identify a local neighborhood appropriate for the system.

\begin{figure}
  \centering
  \includegraphics[width=0.8\linewidth]{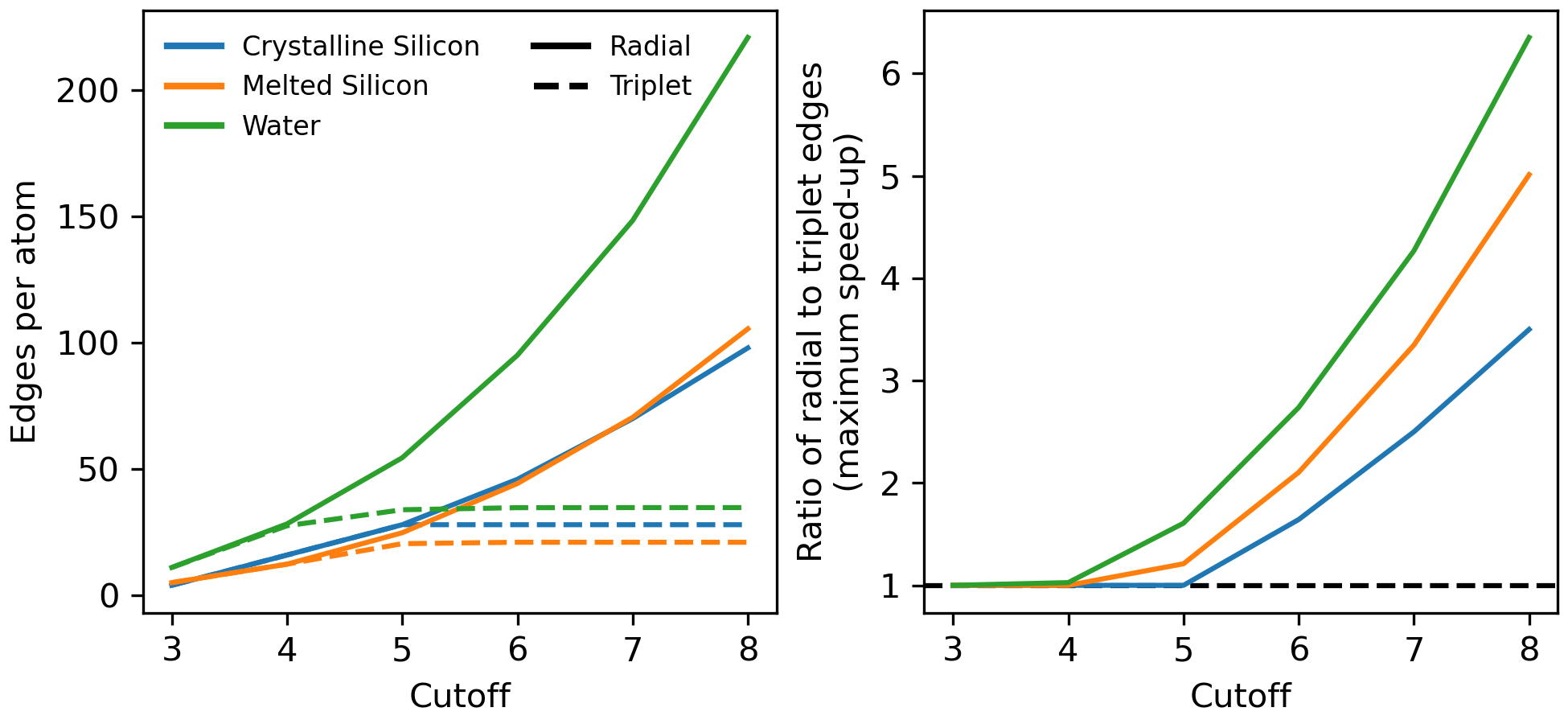}
  \caption{(left) The edges per atom for a radial (solid) envelope function compared to a triplet (dashed) envelope function for crystalline silicon, melted silicon, and liquid water. (right) The ratio of these two are used to find the maximum speed-up for each system.}
  \label{fig:speedup}
\end{figure}

The reduction in the number of edges per atom and the potential speed-up is quantified in Figure \ref{fig:speedup} for different radial cutoffs. 
For radial cutoffs, the number of edges per atom increases cubically with increasing cutoff.
However, triplet screening reduces the number of edges to a constant value starting at 5\AA\ for each system. 
The similarity in the edges between the systems makes it much easier to predict the memory usage of a system for predictably provisioning compute resources and taking advantage of new hot-swapping algorithms like Torch-Sim \cite{cohen2025torchsimefficientatomisticsimulation}. 
Furthermore, as the edges per atom plateau, it becomes possible to increase the radial cutoff significantly to capture longer-range interactions in less dense materials.

The ratio of the radial and triplet edges give the maximum speed-up associated with reducing the edges. 
The dramatic difference in speed-up emphasizes why MLIPs used different cutoffs for different systems before general purpose MLIPs. 
The added cost of extending to a 6\AA\ cutoff for molecular systems often was not worth the additional cost.
However, since the edges have plateaued, triplet envelope functions recover the efficiency of smaller cutoffs in a systematic manner for systems where it is appropriate.

\subsection{Accuracy}
\begin{figure}
  \centering
  \includegraphics[width=\linewidth]{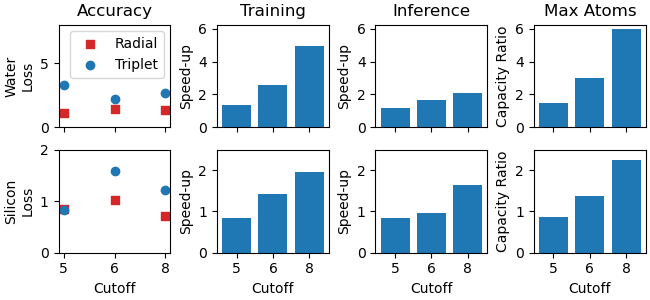}
  \caption{The change in training accuracy, training speed, inference speed, and inference memory pressure when including triplet envelope functions as compared to radial envelope functions. The training speed-up, inference speed-up, and the capacity ratio are the ratio of triplet to radial envelope function showing the improvement of the triplet envelope function in throughput and capacity.}
  \label{fig:results}
\end{figure}

The significant reduction in edges has the potential to adversely impact the accuracy.
To ascertain the effect of removing edges, we perform an ablation study where we train MLIPs with three different radial cutoffs with and without the triplet envelope function on the silicon and water datasets.
The ablation study tests whether triplet sparsification adversely impacts the accuracy of the model.  
The training accuracy is plotted for the three different cutoffs in Figure \ref{fig:results}, showing that there is negligible impact on accuracy, less than previously reported for KNN sparsification of a factor of 2 worse accuracy \cite{rhodes2025orbv3atomisticsimulationscale}. 
The hypothesis that little information is carried in the removed edges is therefore validated.
This aligns with the experience in empirical interatomic potentials, where the receptive field can be restricted to the first few nearest neighbor shells.
Moreover, what is critical is that the models are accurate not just for semiconductor materials like silicon that are more similar to applications of previous MEAM potentials, but also for water, where there are a combination of covalent and van der Waals bonds. 

\subsection{Efficiency}
The benefit of reducing graph edges is geared primarily towards improving the efficiency of MLIPs. 
Both the speed and memory footprint of MLIPs are directly correlated to the number of edges in an atomic graph.
Since there is a cost associated with computing the triplet envelope function, the speed-up has to first compensate for the overhead.
The efficiency is compared for both training and inference, where the training takes advantage of precomputing the edge reduction, which has to be done each evaluation during inference.
By precomputing the edge reduction, the training speed-up approaches 90\% of the ideal speed-up.
For the common cutoff of 6\AA, this translates to doubling the training speed. 
Furthermore, cutoffs that have previously been prohibitively expensive may now be accessible.
We show that at 8\AA\ triplet envelope functions increase training speed by 400\% compared to radial cutoffs.
This opens up avenues to increasing the cutoff of general purpose models to improve performance on applications where increased cutoffs may be necessary to capture even longer range local interactions from, for example, metal-organic-frameworks whose pores can be nanometers in diameter.

The overhead for inference is much higher due to having to compute the triplet sparsification at each simulation step.
Although much more modest, the current triplet implementation is still able to double the inference speed for the 6\AA\ cutoff of water.
The overhead for inference therefore only approaches 60\% of the ideal speed-ups motivating further work on efficient algorithms for triplet computation.

Finally, triplet screening increases the memory efficiency of models to access larger simulation sizes with the same capacity.
By systematically increasing the number of atoms per GPU, we determine the memory pressure of each system.
The increase in capacity given by the maximum number of atoms each GPU can hold is shown in the fourth column of Figure \ref{fig:results}.
The capacity ratios for all cutoffs for both systems approach the ideal values predicted in Figure \ref{fig:speedup}, showcasing the marginal memory overhead of computing triplets.

\subsection{Conservation of Energy}

To show that triplet envelopes do not break energy conservation like KNN sparsification, we run NVT simulations of water and silicon.
The water simulation consists of 20 molecules simulated for 2.5 ps with an NVT ensemble at 300K, while the silicon is simulated for 50 ps with an NVT ensemble at 3000K.
In these conditions, both systems become liquid, with the silicon melting within the first 10ps.
The melting is important as the energy discontinuities occur when the neighbor lists change.

\begin{figure}
  \centering
  \includegraphics[width=0.8\linewidth]{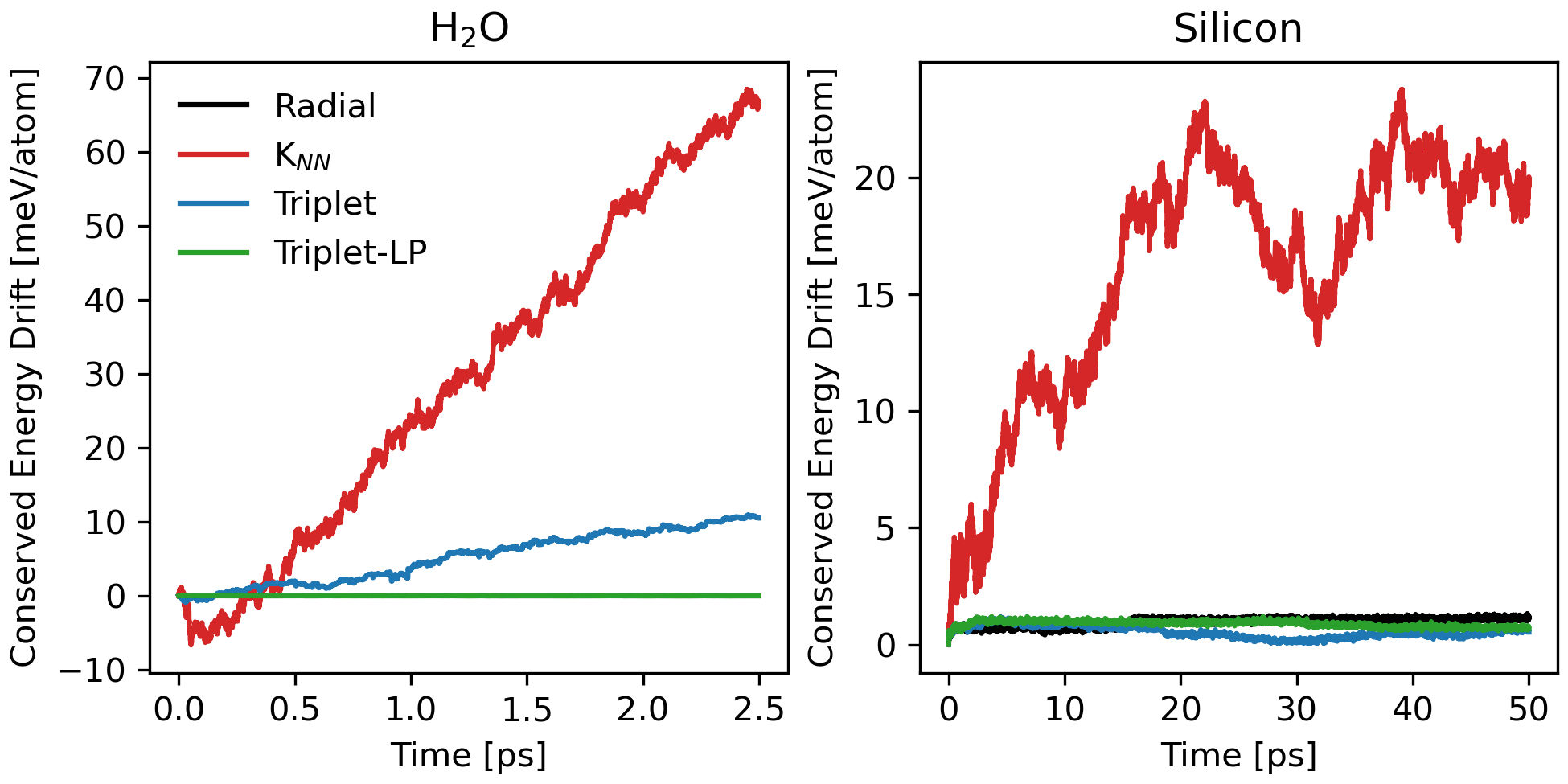}
  \caption{Comparison of the conserved energy of radial, KNN, triplet, and triplet low-pass edge sparsification for models trained on water and silicon.}
  \label{fig:conserved_energy}
\end{figure}

Four different models are used to compare energy conservation of radial, KNN, Triplet, and Triplet with a low-pass filter (Triplet-LP). 
The four models all use a cutoff radius of 6\AA, and the maximum number of neighbors for KNN is set to the common value of 20 \cite{rhodes2025orbv3atomisticsimulationscale}.
The conserved energies for all simulations are shown in Figure \ref{fig:conserved_energy}.
As expected, the radial envelope function conserves energy, but the KNN sparsification does not and the conserved energy fluctuates dramatically.
This is particularly problematic due to recent results showing that energy conservation is a critical factor for accurately computing material observables from the thermodynamic data.
However, since the triplet envelope function smoothly reduces the energy contribution before sparsifying the graph, it is able to conserve energy.

\subsection{Stability}

The final benefit we observed from the triplet envelope function was the increase in simulation stability as reported in Table \ref{tab:stability}.
The stability was assessed by looking at a common failure mode, simulation "explosion".
During the numerical experiments, we recorded any simulations that indicated instability.
Although most simulations were stable, we found that water system with radial cutoffs had a higher propensity for simulation instability.
Showing that ~10\% of simulations were unstable with radial cutoffs, whereas none were unstable with triplet screening.

\begin{table}
 \caption{Stability and data efficiency of radial and triplet envelope functions}
  \centering
  \begin{tabular}{lllll}
     & \multicolumn{2}{c}{Radial Envelope}     & \multicolumn{2}{c}{Triplet Envelope}                 \\
    & Stability     & Dataset Size     & Stability     & Dataset Size  \\
    \midrule
    Silicon & 100\%  & 21  & 100\%  & 21     \\
    Water & 89\%  & 75  & 100\%  & 78     \\
    \bottomrule
  \end{tabular}
  \label{tab:stability}
\end{table}

\section{Discussion and Conclusions}

In this paper we showcased a new type of envelope function based on triplet interactions.
By testing the triplet envelope function on crystalline and liquid systems, we showcased an improvement in training and inference speed as well as a decrease in memory usage, all while impacting the training loss less than previous works on KNN sparsification. 
Furthermore, triplet envelope functions provide smooth potential energy surfaces as evidenced by their conservative dynamics to significantly outperform KNN sparsification.

By combining the results of the loss and stability from Figure \ref{fig:results} and Table \ref{tab:stability}, we reaffirm other studies showing that the training loss is not sufficient for determining the stability of an MLIP \cite{fu2023forcesenoughbenchmarkcritical}.
A critical feature of the increased stability may be that a significant portion of an MLIP is dedicated to learning not just what features contribute to the energy and forces, but also  what features do not contribute and must be learned to ignored.
We test this assertion implicitly through the triplet envelope function that adds an inductive bias that removes a sizable portion of information from the MLIP and may justify exploring methods of reducing the parameter count that may increase performance even more.

As higher order envelope functions become integrated into training and inference pipelines, it will be critical to explore the design space of envelope functions as well as more efficient CPU and GPU algorithms for computing them. 
While the MEAM screening function is a convenient starting point, it is likely not the optimal envelope function as evidenced by the need for low-pass filtering for water. 
Beyond accuracy, triplet functions inspired by other bond-order functionals may better lend themselves to algorithmic improvements.
The $\mathcal{O}(n^3)$ complexity of the MEAM screening function adds significant overhead when the number of neighbors increases at large radial cutoffs. 
Efficient CPU and/or GPU algorithms can aim to address the large overheads associated with computing all triplet interactions yielding a constant cost associated with increasing radial cutoff.

In conclusion, the proposed triplet envelope function offers an exciting new direction to improving the tradeoff between accuracy and cost in MLIP simulations. 
The introduced local, geometric triplet envelope functions enable doubling training and inference speed, tripling memory efficiency, and increasing simulation stability for the most common cutoff of 6\AA.
These multifaceted benefits indicate the promise of creating a new line of research into higher order envelope functions to further engineer MLIP performance.
Importantly, we show that as performance of higher order envelope functions improve, larger cutoff radii above 6\AA (as have been tested in the Allegro MLIP architecture \cite{tan2025highperformancetraininginferencedeep}) may be more widely adopted to capture local effects of open structures. 

\bibliographystyle{unsrt}  
\bibliography{references}

\end{document}